\documentclass[aps,floats,twocolumn]{revtex4}
\usepackage{tabularx,graphicx}
\usepackage{epsfig}
\usepackage{amsmath}
\usepackage{bbm}
\usepackage{graphicx}

\begin{document}

\title{Electron self-energy effects on chiral symmetry breaking in graphene}
\author{J. Gonz\'{a}lez \\}
\address{Instituto de Estructura de la Materia,
        Consejo Superior de Investigaciones Cient\'{\i}ficas, Serrano 123,
        28006 Madrid, Spain}

\date{\today}

\begin{abstract}
We investigate the dynamical breakdown of the chiral symmetry in the theory of Dirac fermions 
in graphene with long-range Coulomb interaction. We analyze the electron-hole 
vertex relevant for the dynamical gap generation in the ladder approximation, showing 
that it blows up at a critical value $\alpha_c$ in the graphene fine structure constant 
which is quite sensitive to many-body corrections. Under static RPA 
screening of the interaction potential, we find that taking into account 
electron self-energy corrections to the vertex increases the critical coupling 
to $\alpha_c \approx 4.9$, for a number $N = 4$ of two-component Dirac 
fermions. When dynamical screening of the interaction is instead considered,
the effect of Fermi velocity renormalization in the electron and hole states 
leads to the value $\alpha_c \approx 1.75$ for $N = 4$, substantially larger 
than that obtained without electron self-energy corrections ($\approx 0.99$), 
but still below the nominal value of the interaction coupling in isolated 
free-standing graphene.

\end{abstract}

\maketitle



\section{Introduction}

The discovery of graphene, the material made of a one-atom-thick carbon layer, 
has attracted a lot attention as it provides the realization of a system
where the electrons have conical valence and conduction bands, therefore
behaving at low energies as massless Dirac fermions\cite{geim,kim,rmp}. 
This offers the possibility of employing the new material as a test ground 
of fundamental concepts in theoretical physics, since the interacting electron
system represents a variant of strongly coupled quantum electrodynamics (QED)
affording quite unusual effects\cite{nil,fog,shy,ter}.

A remarkable feature of such a theory is that a sufficiently strong Coulomb 
interaction may open a gap in the electronic spectrum. 
This effect was already known from the study of QED \cite{appel}, where it
corresponds to the dynamical breakdown of the chiral $U(2)$ symmetry of the
theory. In the context of graphene, such a mechanism is sometimes alluded 
as an exciton instability though, given the absence of a gap between valence
and conduction bands, it becomes more appropriate to describe the effect as
a kind of charge-density-wave instability of the 2D layer. The gap generation
proceeds actually through the development of a non-vanishing average value
of the staggered (sublattice odd) charge density in the underlying honeycomb
lattice, which leads to the generation of a mass and opening of a gap for the 
Dirac quasiparticles.

The question of the dynamical gap generation was first 
addressed in graphene in the approach to the theory with a large number $N$ of 
fermion flavors\cite{khves,gus,ale,son}. The existence of a critical point 
for the formation of an excitonic insulator has been also suggested from 
second-order calculations of electron self-energy corrections\cite{vafek}.
More recently, Monte Carlo simulations of the field theory have been carried 
out in the graphene lattice\cite{drut1,hands}, showing that the chiral symmetry 
of the massless theory can be broken above a critical value for the graphene
fine structure constant
$\alpha_c \approx 1.08$ \cite{drut1}. The possibility of
dynamical gap generation has been also studied in the ladder 
approximation\cite{gama,fer,me,brey}, leading in the case of static screening of 
the interaction to an estimate of the critical coupling 
$\alpha_c \approx 1.62$ for $N = 4$ \cite{gama}. Lately, the resolution 
of the Schwinger-Dyson formulation of the gap equation has revealed that
the effect of the dynamical polarization can 
significantly lower the critical coupling for dynamical gap generation, 
down to a value $\alpha_c \approx 0.92$ for $N = 4$\cite{ggg}.

In this paper we take advantage of the renormalization properties of the 
Dirac theory in order to assess the effect of the electron self-energy 
corrections on the chiral symmetry breaking. In this respect, it has been 
found that the renormalization of the quasiparticle properties can have a 
significant impact, mainly through the increase of the Fermi velocity at 
low energies\cite{khves2,sabio}. Then, we will consider the electron-hole 
vertex accounting for the dynamical gap generation in the ladder 
approximation, shown schematically in Fig. \ref{one}, and we will supplement 
it by self-energy corrections to the electron and hole states. This dressing 
of the quasiparticles will have the result of increasing significantly the 
critical coupling at which the chiral symmetry breaking takes place. Thus,
under static RPA screening of the interaction potential in the ladder series,
we will find the critical value $\alpha_c \approx 4.9$ at the physical number
of flavors $N = 4$. In agreement 
with the trend observed in Ref. \cite{ggg}, we will see however that the more
sensible dynamical screening of the interaction has the effect of lowering 
substantially that estimate, down to a value $\alpha_c \approx 1.75$ which is 
below the nominal value of the interaction coupling in isolated free-standing 
graphene.

\section{Ladder approximation for staggered charge density}

We consider the field theory for Dirac quasiparticles in graphene
interacting through the long-range Coulomb potential, with a Hamiltonian
given by 
\begin{eqnarray}
\lefteqn{H = i v_F \int d^2 r \; \overline{\psi}_i({\bf r}) 
 \boldsymbol{\gamma }   \cdot \boldsymbol{\nabla} \psi_i ({\bf r}) }   \nonumber \\
  &   &     + \frac{e^2}{8 \pi} \int d^2 r_1
\int d^2 r_2 \; \rho ({\bf r}_1) 
       \frac{1}{|{\bf r}_1 - {\bf r}_2|} \rho ({\bf r}_2)  \;\;\;\;\;
\label{ham}
\end{eqnarray}
where  $\{ \psi_i \}$ is a collection of $N/2$ four-component Dirac 
spinors, $\overline{\psi}_i = \psi_i^{\dagger} \gamma_0 $, and 
$\rho ({\bf r}) = \overline{\psi}_i ({\bf r}) \gamma_0 \psi_i ({\bf r})$.
The matrices $\gamma_{\sigma } $ satisfy 
$\{ \gamma_\mu, \gamma_\nu \} = 2 \: {\rm diag } (1,-1,-1)$
and can be conveniently represented in terms of Pauli matrices as
$\gamma_{0,1,2} = (\sigma_3, \sigma_3 \sigma_1, \sigma_3 \sigma_2) \otimes
 \sigma_3$, where the first factor acts on the two sublattice components of 
the graphene lattice. 

Our main interest is to study the behavior of the vertex for the staggered
(sublattice odd) charge density
\begin{equation}
\rho_m ({\bf r}) =  \overline{\psi}({\bf r}) \psi ({\bf r})
\end{equation}
This operator gives the order parameter for the dynamical gap generation, and 
the signal that it gets a nonvanishing expectation value can be obtained from 
the divergence of the response function
$\langle T \rho_m ({\bf q}, t) \rho_m (-{\bf q}, 0) \rangle$.
The singular behavior of this susceptibility can be traced back to the 
divergence at ${\bf q},\omega_q \rightarrow 0$ of the irreducible vertex
\begin{eqnarray}
\lefteqn{\Gamma ({\bf q},\omega_q;{\bf k},\omega_k)  }       \nonumber     \\
  &  &   =  \langle \rho_m ({\bf q},\omega_q) \psi ({\bf k}+{\bf q},\omega_k+\omega_q) 
  \psi^{\dagger} ({\bf k},\omega_k) \rangle_{\rm 1PI}
\end{eqnarray}
where ${\rm 1PI}$ denotes that $\Gamma$ is made of one-particle irreducible diagrams
without external electron propagators.

\begin{figure}
\begin{center}
\mbox{\epsfxsize 9.0cm \epsfbox{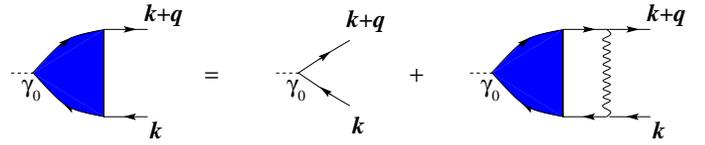}}
\end{center}
\caption{Self-consistent diagrammatic equation for the vertex
$\langle \rho_m ({\bf q},\omega_q) \psi ({\bf k}+{\bf q},\omega_k+\omega_q) 
  \psi^{\dagger} ({\bf k},\omega_k) \rangle$, 
equivalent to the sum of ladder diagrams built from
the iteration of the Coulomb interaction (wavy line) between electron and
hole states (arrow lines).}
\label{one}
\end{figure}

In the ladder approximation, the vertex $\Gamma $ is bound to satisfy the 
self-consistent equation depicted diagrammatically in Fig. \ref{one}. This 
equation can be solved perturbatively by iterating the interaction between 
electrons and holes in the vertex, in which case this ends up being represented  
by the sum of ladder diagrams. On the other hand, the self-consistent 
equation can be written in compact form, specially at momentum transfer
${\bf q} = 0$ and $\omega_q = 0$. We recall at this point the expression of 
the free Dirac propagator
\begin{equation}
\langle \psi ({\bf k}, \omega_k )  \psi^{\dagger } ({\bf k}, \omega_k ) \rangle_{\rm free} 
  =  i \frac{-\gamma_0 \omega_k  + v_F \boldsymbol{\gamma} \cdot {\bf k} }
                      {-\omega_k^2  + v_F^2 {\bf k}^2 - i\eta } \gamma_0
\end{equation}
Given that $\Gamma $ must be anyhow proportional to $\gamma_0 $, we get
\begin{align}
 & -\frac{-\gamma_0 \omega_p  + v_F \boldsymbol{\gamma} \cdot {\bf p} }
                      {-\omega_p^2  + v_F^2 {\bf p}^2 - i\eta } 
   \gamma_0  \: \Gamma ({\bf 0},0;{\bf p},\omega_p) \:
\frac{-\gamma_0 \omega_p  + v_F \boldsymbol{\gamma} \cdot {\bf p} }
          {-\omega_p^2  + v_F^2 {\bf p}^2 - i\eta } \gamma_0    \nonumber  \\
 & = \frac{\Gamma ({\bf 0},0;{\bf p},\omega_p)}{-\omega_p^2  + v_F^2 {\bf p}^2 - i\eta } 
\end{align}
The self-consistent equation for the vertex becomes then
\begin{eqnarray}
\lefteqn{\Gamma ({\bf 0},0;{\bf k},i\omega_k) = \gamma_0  }    \nonumber      \\ 
 &  &     +  \int \frac{d^2 p}{(2\pi )^2} \frac{d\omega_p}{2\pi } 
       \frac{\Gamma ({\bf 0},0;{\bf p},i\omega_p)}{\omega_p^2 + v_F^2{\bf p}^2} 
               V({\bf k}-{\bf p},i\omega_k - i\omega_p)   \;\;\;\;\;
\label{self}
\end{eqnarray}
where $V({\bf p},\omega_p)$ stands for the Coulomb interaction. We will deal
in general with the RPA to screen the potential, so that 
\begin{equation}
V({\bf p}, \omega_p) = \frac{e^2}{2 |{\bf p}| + e^2  \chi ({\bf p}, \omega_p)}
\end{equation}
in terms of the polarization $\chi $ for $N$ two-component Dirac fermions.

Eq. (\ref{self}) is formally invariant under a 
dilatation of frequencies and momenta, which shows that the scale
of $\Gamma ({\bf 0},0;{\bf k},\omega_k)$ is dictated by the high-energy cutoff 
$\Lambda $ needed to regularize the integrals. The vertex acquires in general 
an anomalous dimension $\gamma_{\psi^2}$, which governs the behavior under 
changes in the energy scale\cite{amit} 
\begin{equation}
\Gamma ({\bf q},\omega_q;{\bf k},\omega_k) \sim \Lambda^{\gamma_{\psi^2}}
\end{equation}
We recall below how to compute $\gamma_{\psi^2}$, showing that it diverges 
at a critical value of the interaction strength $\alpha = e^2/4\pi v_F$. 
This translates into a divergence of the own susceptibility 
$\langle T \rho_m ({\bf q}, t) \rho_m (-{\bf q}, 0) \rangle$
at momentum transfer ${\bf q} \rightarrow 0$, providing then the signature
of the condensation of 
$\rho_m ({\bf r}) =  \overline{\psi}({\bf r}) \psi ({\bf r})$
and the consequent development of the gap for the Dirac quasiparticles.

\section{Electron self-energy effects in statically screened ladder approximation}

We deal first with the approach in which electrons and holes are dressed by 
self-energy corrections, while the Coulomb interaction in (\ref{self}) is 
screened by means of the static RPA with polarization 
\begin{equation}
\chi ({\bf p}, 0) = \frac{N}{16}\frac{|{\bf p}|}{v_F}
\end{equation}
The most important self-energy effect comes from the renormalization of the 
Fermi velocity at low energies\cite{np2,prbr}, which can be incorporated by
replacing $v_F$ in Eq. (\ref{self}) by the effective Fermi velocity
\begin{equation}
\widetilde{v}_F({\bf p})  =   v_F   + \Sigma_v ({\bf p})
\end{equation} 
dressed with the 
self-energy corrections $\Sigma_v ({\bf p})$. The expansion of Eq. (\ref{self})
in powers of $\Sigma_v ({\bf p})$ would amount to the iteration of
self-energy corrections in the electron and hole internal lines 
in Fig. \ref{one}, showing that the present approach encodes a systematic 
way of improving the sum of ladder diagrams for the vertex $\Gamma $ \cite{note}.

The electron self-energy corrections, as well as the terms of the 
ladder series, are given by logarithmically divergent integrals that need to
be cut off at a high-energy scale $\Lambda $. Alternatively, one can also define
the theory at spatial dimension $D = 2 - \epsilon$, what automatically 
regularizes all the momentum integrals. After performing the frequency integral, 
Eq. (\ref{self}) then becomes
\begin{eqnarray}
\lefteqn{\Gamma ({\bf 0},0;{\bf k},\omega_k) = \gamma_0  }    \nonumber    \\
  &  &    +  \frac{e_0^2}{4\kappa } 
   \int \frac{d^D p}{(2\pi )^D} \Gamma ({\bf 0},0;{\bf p},\omega_k) 
    \frac{1}{\widetilde{v}_F({\bf p}) |{\bf p}|} \frac{1}{|{\bf k}-{\bf p}|}
\label{selfcons}
\end{eqnarray}
where $e_0^2$ is related to $e^2$ through an auxiliary 
momentum scale $\rho $ such that 
\begin{equation}
e_0^2 = \rho^{\epsilon} e^2 
\end{equation}
and we have defined the dielectric constant
\begin{equation}
\kappa = 1 + \frac{N e^2}{32 v_F}
\end{equation}

\begin{figure}
\begin{center}
\mbox{\epsfxsize 3.0cm \epsfbox{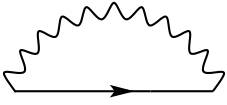}}
\end{center}
\caption{Electron self-energy correction leading to a divergent renormalization 
of the Fermi velocity $v_F$.}
\label{rainbow}
\end{figure}

In the ladder approximation, the Fermi velocity gets a divergent correction
only from the ^^ ^^ rainbow'' self-energy diagram with exchange of a single 
screened interaction shown in Fig. \ref{rainbow} \cite{np2}. The dressed Fermi 
velocity becomes
\begin{equation}
\widetilde{v}_F({\bf p}) = v_F + \frac{e_0^2}{16 \pi^2\kappa} 
   (4\pi )^{\epsilon /2}   
  \frac{\Gamma (\tfrac{\epsilon }{2}) \Gamma (\tfrac{1-\epsilon}{2}) 
                                            \Gamma (\tfrac{3-\epsilon}{2}) }
  {  \Gamma (2 - \epsilon) }
   \frac{1}{|{\bf p}|^{\epsilon}}
\label{vd}
\end{equation}
The expressions (\ref{selfcons}) and (\ref{vd}) are singular in the limit 
$\epsilon \rightarrow 0$. The most convenient way to show that all the poles 
in the $\epsilon $ parameter can be renormalized away is to resort at this point 
to a perturbative computation of $\Gamma ({\bf 0},0;{\bf k},\omega_k)$.

The solution of (\ref{selfcons}) can be obtained in the form
\begin{equation}
\Gamma ({\bf 0},0;{\bf k},\omega_k) = 
 \gamma_0 
 \left(1 + \sum_{n=1}^{\infty} \lambda_0^n 
                           \frac{r_n }{|{\bf k}|^{n\epsilon}} \right)
\end{equation}
with $\lambda_0 = e_0^2/4\pi \kappa v_F $.
Each term in the sum can be obtained from the previous one by expanding
$1/\widetilde{v}_F({\bf p})$ in Eq. (\ref{selfcons}) in powers of $e_0^2$ 
and noticing that
\begin{eqnarray}
\lefteqn{ \int \frac{d^D p}{(2\pi )^D} \frac{1}{|{\bf p}|^{(m-1)\epsilon} }
            \frac{1}{|{\bf p}|} \frac{1}{|{\bf k}-{\bf p}|}            }
                                                            \nonumber   \\
  & &  
 =  \frac{(4\pi )^{\epsilon /2}}{4 \pi^{3/2}}   
  \frac{\Gamma \left(\tfrac{m\epsilon }{2} \right) \Gamma \left(\tfrac{1-m\epsilon}{2} \right) 
                                \Gamma \left(\tfrac{1-\epsilon}{2} \right) }
  { \Gamma \left(\tfrac{1+(m-1)\epsilon}{2} \right) \Gamma \left(1-\tfrac{m + 1}{2}\epsilon \right) }
 \frac{1}{|{\bf k}|^{m\epsilon}}  
\end{eqnarray}
At each given perturbative level, the vertex displays then a number of poles 
at $\epsilon = 0$. The crucial point is that these divergences can be
reabsorbed by passing to physical quantities defined by the multiplicative
renormalization
\begin{eqnarray} 
   v_F  & = &   Z_v(v_F)_{\rm ren}                                    \\       
\overline{\psi} \psi  & = &   Z_{\psi^2} (\overline{\psi} \psi )_{\rm ren}
\end{eqnarray}
We observe that the scale of the single Dirac field $\psi $ does not need to be 
renormalized in this approach, as self-energy corrections of the 
form shown in Fig. \ref{rainbow} with a statically screened interaction do not 
modify the frequency dependence of the Dirac propagator.

The renormalized vertex
\begin{equation}
\Gamma_{\rm ren} = Z_{\psi^2} \Gamma
\end{equation}
can be actually made finite at $\epsilon = 0$ with a 
suitable choice of momentum-independent factors $Z_v$ and $Z_{\psi^2}$. 
$Z_v$ must be chosen to cancel the $1/\epsilon $ pole arising from 
$\Gamma (\epsilon /2 )$ in (\ref{vd}), and it 
has therefore the simple structure 
\begin{equation}
Z_v = 1 +  \frac{b_1}{\epsilon } 
\end{equation}
with $b_1 = - e^2/16\pi \kappa (v_F)_{\rm ren}$. On the other hand, we have 
the general structure
\begin{equation}
Z_{\psi^2} = 1 + \sum_{i=1}^{\infty} \frac{c_i }{\epsilon^i}
\label{poles}
\end{equation}
The position of the different poles must be chosen to enforce the
finiteness of $\Gamma_{\rm ren} = Z_{\psi^2} \Gamma $ in the limit
$\epsilon \rightarrow 0$. The computation of the 
first orders of the expansion gives for instance the result
\begin{eqnarray}
c_1 (\lambda ) & = &  - \frac{1}{2} \lambda - \tfrac{1}{8} \log(2) \: \lambda^2 
      - \tfrac{1}{1152} \left( \pi ^2 + 120 \log ^2(2) \right)   \lambda^3     \nonumber   \\
  &  &   -  \tfrac{10 \pi ^2 \log (2)+688 \log ^3(2)+15 \zeta (3)}{6144}   \lambda^4  
                                                               \nonumber  \\
 & & - \tfrac{13 \pi ^4+2064 \pi ^2 \log ^2(2)+144 \left(716 \log ^4(2)+37 \log (2) \zeta (3)\right)}{737280} \: \lambda^5    
                                                                             \nonumber \\
  &  &                                                +    \ldots             \nonumber \\
c_2 (\lambda ) & = &  \tfrac{1}{16} \: \lambda^2 + 
 \tfrac{1}{24} \log(2) \: \lambda^3                                            \nonumber  \\
   &  & + \tfrac{1}{18432} \left( 5 \pi ^2 + 744 \log ^2(2) \right)  \lambda^4  \nonumber  \\
   &  &   + \tfrac{110 \pi ^2 \log (2)+8592 \log ^3(2)+135 \zeta (3)}{184320}  \: \lambda^5 
                                                         +  \ldots         \nonumber  \\
c_3 (\lambda ) & = & - \tfrac{1}{768} \log (2) \: \lambda^4 - \tfrac{1}{184320} \left( \pi ^2+360 \log ^2(2) \right)  \lambda^5 
                                                                            \nonumber  \\
    &  &                                                          + \ldots   \nonumber  \\
c_4 (\lambda ) & = & - \tfrac{1}{7680} \log (2) \: \lambda^5  + \ldots  
\label{coeff}
\end{eqnarray}
where the series are written in terms of the renormalized coupling
$\lambda$ defined by 
\begin{equation}
\lambda \equiv \rho^{-\epsilon} Z_v \lambda_0 = \frac{e^2}{4\pi \kappa (v_F)_{\rm ren}}
\end{equation}

The physical observable in which we are interested is the anomalous 
dimension $\gamma_{\psi^2}$. The change in the dimension of $\Gamma_{\rm ren}$
comes from the dependence of $Z_{\psi^2}$ on the only dimensionful scale 
$\rho $ in the renormalized theory. Therefore we have\cite{amit} 
\begin{equation}
\gamma_{\psi^2} = \frac{\rho }{Z_{\psi^2}} \frac{\partial Z_{\psi^2} }{\partial \rho }
\end{equation}
The original bare theory at $D \neq 2$ does not know about the arbitrary scale 
$\rho $, and the 
independence of $\lambda_0 = \rho^{\epsilon} \lambda /Z_v $ on that
variable leads to 
\begin{equation}
\rho \frac{\partial \lambda }{\partial \rho } = 
 - \epsilon \lambda - \lambda b_1 (\lambda )
\label{rge}
\end{equation}
At $\epsilon = 0$, this is the well-known expression of the scale 
dependence of the effective interaction strength, arising from the 
renormalization of the Fermi velocity\cite{np2}.
The anomalous dimension becomes finally\cite{ram}
\begin{equation}
\gamma_{\psi^2} = \frac{\rho }{Z_{\psi^2}}
  \frac{\partial \lambda }{\partial \rho }
  \frac{\partial  Z_{\psi^2} }{\partial \lambda }
 = - \lambda \frac{d c_1}{d \lambda }
\label{dreg}
\end{equation}

In the derivation of (\ref{dreg}), it is implicitly assumed that poles 
in the $\epsilon $ parameter cannot appear at the right-hand-side of the 
equation. For this to be true, the set of equations
\begin{equation} 
\frac{d c_{i+1}}{d \lambda } = c_i \frac{d c_1}{d \lambda } - b_1 \frac{d c_i}{d \lambda } 
\end{equation}
must be satisfied identically\cite{ram}. 
Quite remarkably, we have verified that this is the case, up to the order 
$\lambda^{17} $ we have been able to compute numerically the coefficients in 
(\ref{poles}). This is the proof of
the renormalizability of the theory, which guarantees that physical quantities
like $\gamma_{\psi^2}$ remain finite in the limit $\epsilon \rightarrow 0$.

\begin{figure}
\begin{center}
\mbox{\epsfxsize 7.0cm \epsfbox{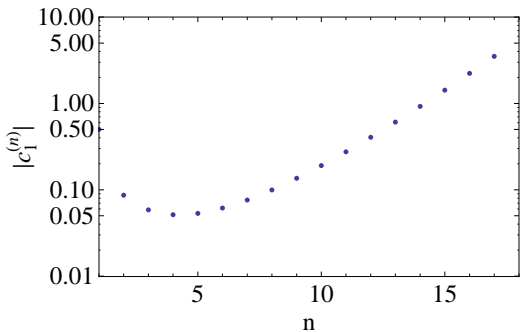}}
\end{center}
\caption{Plot of the absolute value of the coefficients $c_1^{(n)}$ in the 
expansion of $c_1 (\lambda )$ as a power series of the coupling $\lambda $.}
\label{c1n}
\end{figure}

From the practical point of view, the important result is the evidence that 
the perturbative expansion of $c_1 (\lambda )$ 
\begin{equation}
c_1 (\lambda ) = \sum_{n} c_1^{(n)} \lambda^n
\end{equation}
approaches a geometric series 
in the $\lambda $ variable. The plot of the coefficients $c_1^{(n)}$ 
computed numerically up to order $\lambda^{17} $ is shown in Fig.
\ref{c1n}. It can be checked that the coefficients
grow exponentially with the order $n$, in such a way that
\begin{equation}
- c_1 (\lambda ) \geq \sum_{n=1}^{\infty} d^n \lambda^n \; + \; {\rm regular}  \;\;\; {\rm terms}
\end{equation}

\begin{figure}
\begin{center}
\mbox{\epsfxsize 6.0cm \epsfbox{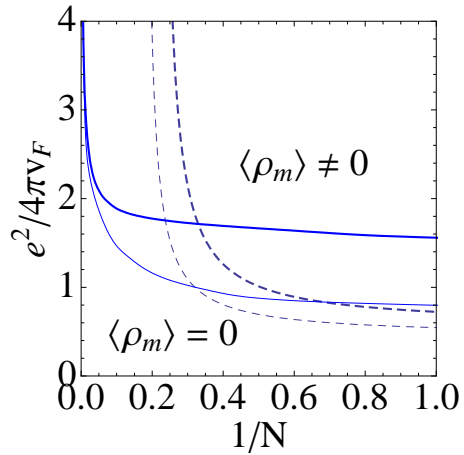}}
\end{center}
\caption{Phase diagram showing the boundary between the metallic phase and
the phase with dynamical gap generation ($\langle \rho_m \rangle \neq 0$) 
in the ladder approximation.
The thin dashed (solid) line represents the phase boundary obtained with static
(dynamic) RPA screening of the interaction potential and no electron self-energy 
corrections. The thick dashed (solid) line represents the boundary after including
the effect of the electron self-energy corrections on top of the static (dynamic) 
RPA screening of the interaction in the ladder series.
}
\label{two}
\end{figure}

An estimate of $d$ can be obtained from the coefficients available in the perturbative 
series of $c_1 (\lambda )$. The ratio between consecutive $c_1^{(n)}$ increases with 
the order $n$, converging towards a limit value.
The best fit of the asymptotic behavior allows us to estimate a radius of convergence 
\begin{equation}
\lambda_c \approx 0.56
\end{equation}
This has to be compared with the value 
found in the approach neglecting self-energy corrections, which leads to 
$\lambda_c \approx 0.45$ \cite{me}, in close agreement with the result of 
Ref. \onlinecite{gama}. The critical coupling in the variable 
$\lambda $ can be used to draw the boundary for dynamical 
gap generation in the space of $N$ and $\alpha = e^2/4\pi (v_F)_{\rm ren} $, 
recalling that
\begin{equation}
\lambda  =  \frac{\alpha}{1 + \frac{N \pi}{8} \alpha }
\end{equation}
The corresponding phase diagram is represented in Fig. \ref{two}.
For $N = 4$, we get in particular the critical coupling $\alpha_c \approx 4.9$,
significantly above the critical value that would be obtained from the radius
of convergence without self-energy corrections ($\alpha_c \approx 1.53$).

\section{Electron self-energy effects in dynamically screened ladder approximation}

In the framework of the ladder approximation, one can also study the effect
of electron self-energy corrections under dynamical screening of the Coulomb 
interaction potential. We can improve
the static RPA by considering the full effect of the 
frequency-dependent polarization, which for Dirac fermions 
takes the form\cite{np2}
\begin{equation}
\chi ({\bf p}, \omega_p) = 
\frac{N}{16} \frac{{\bf p}^2}{\sqrt{v_F^2 {\bf p}^2 - \omega_p^2}}
\label{dyn}
\end{equation} 
This expression can be introduced in Eq. (\ref{self}) to look again for
self-consistent solutions for the vertex $\Gamma ({\bf 0},0;{\bf k},\omega_k)$. 
Given that in this case we must resort to numerical methods for the
resolution of the integral equation, we can go beyond the self-energy effects 
considered before by taking into account the electron self-energy corrections 
in the RPA improved with the polarization (\ref{dyn}). In this approach, 
the behavior of the dressed Fermi velocity $\widetilde{v}_F({\bf p})$ 
is given as a function of $g = N e^2/32 \widetilde{v}_F$ by the nonlinear 
equation\cite{prbr}
\begin{equation}
 \frac{\partial \log \widetilde{v}_F}{\partial \log |{\bf p}|}  = 
 - \frac{8}{N\pi^2} 
\left( 1 +  \frac{\arccos g}{g \sqrt{1-g^2}} - \frac{\pi}{2} \frac{1}{g}
\right)   
\label{vflow}
\end{equation}
We have then used the solution of Eq. (\ref{vflow}) to replace $v_F$ in
Eq. (\ref{self}) by the momentum dependent $\widetilde{v}_F$, which
represents a significant improvement in the sum of self-energy corrections
in the ladder series for the vertex.

In this procedure, we find again that there is a critical coupling in the
variable $\alpha = e^2/4\pi v_F$ at which $\Gamma ({\bf 0},0;{\bf k},\omega_k)$ 
blows up, marking the boundary between two different regimes where 
Eq. (\ref{self}) has respectively positive and negative solutions.
In practice, we have solved the integral equation by defining the
vertex in a discrete set of points in frequency and momentum space. One can 
take as independent variables in $\Gamma ({\bf 0},0;{\bf k},\omega_k)$ the modulus of 
${\bf k}$ and positive frequencies $\omega_k $. We have adopted accordingly 
a grid of dimension $l \times l$ covering those variables, with 
$l$ running up to a value of 200 for which it is still viable to invert a 
matrix of dimension $l^2$.

As a check of our approach, we have compared the results of the numerical
diagonalization of (\ref{self}), still keeping the undressed Fermi velocity
$v_F$, with the values of the critical coupling in Ref. \cite{ggg}, where the 
resolution of the gap equation has been accomplished with the frequency-dependent 
polarization. We have relied on the scale invariance of our model to find the 
trend of $\alpha_c$ at large $l$, as the critical coupling must obey a 
finite-size scaling law 
\begin{equation}
\alpha_c (l) = \alpha_c (\infty ) + \frac{c}{l^\nu }
\end{equation}
At $N = 4$, we get $\alpha_c (200) \approx 1.08$ and the estimate 
$\alpha_c (\infty ) \approx 0.99$, which turns out to be close to the critical
coupling $\alpha_c \approx 0.92$ found in Ref. \cite{ggg}, providing a nice 
check of our computational approach in the case of unrenormalized $v_F$.

The electron self-energy corrections lead anyhow to a substantial increase
in the values of the critical coupling $\alpha_c (l)$. This is a decreasing 
function of $l$, as the limit $l \rightarrow \infty$ corresponds to the 
large-volume limit of the system. Then, as a result of diagonalizing 
Eq. (\ref{self}) with the effective $\widetilde{v}_F$, we have chosen to 
represent in Fig. \ref{two} the upper bound $\alpha_c (200)$ to the critical
coupling as a function of $N$. 
We observe that, for $N \lesssim 3$, the values of the 
critical coupling are larger than those obtained with static screening of
the interaction potential, while the situation is inverted for $N \gtrsim 3$.
In coincidence with the findings of Ref. \cite{ggg}, there is indeed no upper 
limit on $N$ for the onset of chiral symmetry breaking in this approach.  
At $N = 4$, we get 
\begin{equation}
\alpha_c (200) \approx 1.75
\end{equation}
which is substantially
smaller than the value found in Sec. III with the static RPA screening in 
the ladder series. These results support the idea that, in the particular
case of graphene ($N = 4$), the nominal coupling of the system in vacuum 
($\alpha \approx 2.2$) should be above the critical coupling for dynamical
gap generation. This is reinforced by the fact that other
effects neglected thus far have to do with the electron
self-energy corrections to the own polarization $\chi$. These should lead
to a reduction of the screening and the consequent enhancement of the 
effective interaction strength. The values that we find for $\alpha_c$ 
should be taken in this regard as an upper bound for the critical coupling, 
at least when compared with the
result of including the effect of Fermi velocity renormalization in the bare 
polarization.

\section{Conclusion}

In this paper we have considered the impact that electron self-energy 
corrections may have on the chiral symmetry breaking in the interacting theory 
of Dirac fermions. Our starting point has been the ladder approximation for 
the electron-hole vertex appearing in the response function for dynamical
gap generation, which we have supplemented by including systematically the 
self-energy corrections to electron and hole states in the 
ladder series.

In this framework, we have been able to account for the effect of the 
Fermi velocity renormalization on the critical coupling for dynamical gap
generation. In this respect, it has been already suggested that 
the growth of the Fermi velocity at low energies can have a deep impact to 
prevent the chiral symmetry breaking\cite{sabio,see}. The scale dependence
of the Fermi velocity, expressed nonperturbatively in Eq. (\ref{vflow}), has 
been already observed in experiments with graphene at very low doping 
levels\cite{paco}. Our results show actually that the effect of 
renormalization of the Fermi velocity induces a significant reduction in the 
strength of the dynamical symmetry breaking in graphene, leading to a critical 
coupling $\alpha_c \approx 4.9$ in the case of static RPA screening of the 
interaction potential in the ladder series, and to a value 
$\alpha_c \approx 1.75$ in the more sensible instance of dynamical screening 
of the interaction.

One of the main conclusions of this work is that the screening effects must be
treated accurately in order to make a reliable estimate of the
critical coupling for dynamical gap generation in graphene. This is so 
as such an instability depends strongly on the singular behavior of 
the Coulomb interaction in the undoped system. In this regard, the situation 
is quite different to the case of bilayer graphene, where several low-energy 
instabilities have been also predicted\cite{vaf,zhang,nan,lemo}. These can
be traced back to the divergence of objects like the electron-hole 
polarization, which results from the particular form of the bandstructure
and does not require a long-range interaction. In monolayer graphene, the
instability towards chiral symmetry breaking appears to be quite sensitive
to many-body corrections to the Coulomb interaction, which makes more delicate
the precise computation of the critical interaction strength.

The other important conclusion is that the value $\alpha_c \approx 1.75$
resulting from the self-energy corrections still remains below the nominal 
coupling for graphene in vacuum. This means that an isolated free-standing
layer of the material should be in the phase with dynamical gap generation, 
which is apparently at odds with present experimental measurements in suspended 
graphene samples. A key observation is however that, if chiral symmetry
breaking is to proceed in graphene according to the present estimates, it is 
going to lead to a gap at least three orders of magnitude below the 
high-energy scale of the Dirac theory, as found in the resolution of the gap 
equation\cite{ggg}. This suggests then that the dynamical gap generation 
cannot be discarded in isolated free-standing graphene, though its 
experimental signature may be only found in suitable samples, for which the 
Fermi level can be tuned within an energy range below the meV scale about
the charge neutrality point.

{\em Acknowledgments.---}
We thank F. Guinea and V. P. Gusynin for very useful discussions.
The financial support from MICINN (Spain) through grant
FIS2008-00124/FIS is also acknowledged.


\begin{thebibliography}{99}




\bibitem{geim}
K. S. Novoselov, A. K. Geim, S. V. Morozov, D. Jiang, M. I. Katsnelson,
I. V. Grigorieva, S. V. Dubonos and A. A. Firsov, Nature {\bf 438}, 197 (2005).

\bibitem{kim}
Y. Zhang, Y.-W. Tan, H. L. Stormer and P. Kim, Nature {\bf 438}, 201 (2005).

\bibitem{rmp}
A. H. Castro Neto, F. Guinea, N. M. R. Peres, K. S. Novoselov and
A. K. Geim, Rev. Mod. Phys. {\bf 81}, 109 (2009).

\bibitem{nil}
V. M. Pereira, J. Nilsson and A. H. Castro Neto, Phys. Rev. Lett. {\bf 99}, 
166802 (2007).

\bibitem{fog}
M. M. Fogler, D. S. Novikov, and B. I. Shklovskii, Phys. Rev. B {\bf 76}, 
233402 (2007).

\bibitem{shy}
A. V. Shytov, M. I. Katsnelson, and L. S. Levitov, 
Phys. Rev. Lett. {\bf 99}, 236801 (2007).


\bibitem{ter}
I. S. Terekhov, A. I. Milstein, V. N. Kotov, and O. P. Sushkov,
Phys. Rev. Lett. {\bf 100}, 076803 (2008).
 

\bibitem{appel}
T. Appelquist, D. Nash and L. C. R. Wijewardhana, Phys. Rev. Lett. {\bf 60},
2575 (1988).

\bibitem{khves}
D. V. Khveshchenko, Phys. Rev. Lett. {\bf 87}, 246802 (2001).

\bibitem{gus}
E. V. Gorbar, V. P. Gusynin, V. A. Miransky and I. A. Shovkovy, Phys. Rev. 
B {\bf 66}, 045108 (2002).

\bibitem{ale}
I. L. Aleiner, D. E. Kharzeev and A. M. Tsvelik, Phys. Rev. B {\bf 76},
195415 (2007).

\bibitem{son}
J. E. Drut and D. T. Son, Phys. Rev. B {\bf 77}, 075115 (2008).

\bibitem{vafek}
O. Vafek and M. J. Case, Phys. Rev. B {\bf 77}, 033410 (2008).  


\bibitem{drut1}
J. E. Drut and T. A. L\"ahde, Phys. Rev. Lett. {\bf 102}, 026802 (2009);
Phys. Rev. B {\bf 79}, 241405(R) (2009).

\bibitem{hands}
See also S. J. Hands and C. G. Strouthos, Phys. Rev. B {\bf 78}, 165423 (2008);
W. Armour, S. Hands, C. Strouthos, Phys. Rev. B {\bf 81}, 125105 (2010).

\bibitem{gama}
O. V. Gamayun, E. V. Gorbar and V. P. Gusynin, Phys. Rev. B {\bf 80},
165429 (2009).

\bibitem{fer}
J. Wang, H. A. Fertig and G. Murthy, Phys. Rev. Lett. {\bf 104}, 186401 (2010).

\bibitem{me}
J. Gonz\'alez, Phys. Rev. B {\bf 82}, 155404 (2010).

\bibitem{brey}
J. Wang, H. A. Fertig, G. Murthy and L. Brey, Phys. Rev. B {\bf 83}, 
035404 (2011).

\bibitem{ggg}
O. V. Gamayun, E. V. Gorbar and V. P. Gusynin, Phys. Rev. B {\bf 81},
075429 (2010).

\bibitem{khves2}
D. V. Khveshchenko, J. Phys.: Condens. Matter {\bf 21}, 075303 (2009).

\bibitem{sabio}
J. Sabio, F. Sols and F. Guinea, Phys. Rev. B {\bf 82}, 121413͑(R) (2010).


\bibitem{amit}
D. J. Amit and V. Mart\'{\i}n-Mayor, {\em Field Theory, the Renormalization 
Group, and Critical Phenomena}, Chaps. 6 and 8 
(World Scientific, Singapore, 2005).

\bibitem{np2}
J. Gonz\'alez, F. Guinea and M. A. H. Vozmediano,
Nucl. Phys. B {\bf 424}, 595 (1994).

\bibitem{prbr}
J. Gonz\'alez, F. Guinea and M. A. H. Vozmediano,
Phys. Rev. B {\bf 59}, R2474 (1999).

\bibitem{note}
We note that the vertex may depend on 
further many-body corrections to the screening of the interaction potential, 
though the most significant effect of Fermi velocity renormalization in
the polarization, not considered here, can only lead to a reduction of
the screening and therefore to lower values of the 
critical coupling for chiral symmetry breaking.

\bibitem{ram}
P. Ramond, {\em Field Theory: A Modern Primer}, Chap. IV (Benjamin/Cummings, Reading, 1981).


\bibitem{see}
See also I. F. Herbut, V. Juri\v{c}i\'c and O. Vafek, Phys. Rev. B {\bf 80}, 075432 (2009); 
V. Juri\v{c}i\'c, I. F. Herbut and G. W. Semenoff, Phys. Rev. B {\bf 80}, 081405 (2009).

\bibitem{paco}
D. C. Elias, R. V. Gorbachev, A. S. Mayorov, S. V. Morozov, A. A. Zhukov, P. Blake,
L. A. Ponomarenko, I. V. Grigorieva, K. S. Novoselov, F. Guinea and A. K. Geim,
Nature Phys. {\bf 7}, 701 (2011).

\bibitem{vaf}
O. Vafek and K. Yang, Phys. Rev. B {\bf 81}, 041401(R) (2010).

\bibitem{zhang}
F. Zhang, H. Min, M. Polini and A. H. MacDonald, Phys. Rev. B {\bf 81}, 
041402(R) (2010).

\bibitem{nan}
R. Nandkishore and L. Levitov, Phys. Rev. Lett. {\bf 104}, 156803 (2010).

\bibitem{lemo}
Y. Lemonik, I. L. Aleiner, C. Toke and V. I. Fal'ko, Phys. Rev. B {\bf 82}, 
201408(R) (2010).





\end{thebibliography}
\end{document}